\documentclass[aps,preprint,prd,eqsecnum,nofootinbib,showpacs,12pt]{revtex4}

\usepackage{amssymb,latexsym} 
\usepackage{epsfig}
\usepackage{bm}

\newcommand{\be}{\begin{equation}} \newcommand{\ee}{\end{equation}}

\newcommand{\ba}{\begin{eqnarray}}
\newcommand{\ea}{\end{eqnarray}}

\newcommand{\cplus}{k_+}
\newcommand{\cminus}{k_-}
\newcommand{\cmixed}{k_0}

\begin{document}
\thispagestyle{empty} 

\date{\today}

\title{
{\vspace{-1.2em} \parbox{\hsize}{\hbox to \hsize 
{\hss \normalsize\rm IFUP-TH 2004/19, UPRF-2004-06}}} \\
Phase diagram of the lattice Wess-Zumino model
from\\ rigorous lower bounds on the energy}

\author{Matteo Beccaria}%
\email{Matteo.Beccaria@le.infn.it}%
\affiliation{INFN, Sezione di Lecce, and
Dipartimento di Fisica dell'Universit\`a di Lecce,
Via Arnesano, ex Collegio Fiorini, I-73100 Lecce, Italy}

\author{Massimo Campostrini}%
\email{Massimo.Campostrini@df.unipi.it}%
\affiliation{INFN, Sezione di Pisa, and
Dipartimento di Fisica ``Enrico Fermi'' dell'Universit\`a di Pisa,
Via Buonarroti 2, I-56125 Pisa, Italy}

\author{Gian Fabrizio De Angelis}%
\email{Gianfabrizio.DeAngelis@le.infn.it}%
\affiliation{INFN, Sezione di Lecce, and
Dipartimento di Fisica dell'Universit\`a di Lecce,
Via Arnesano, ex Collegio Fiorini, I-73100 Lecce, Italy}

\author{Alessandra Feo}%
\email{feo@fis.unipr.it}%
\affiliation{
Dipartimento di Fisica, Universit\`a di Parma and INFN Gruppo Collegato di Parma,
Parco Area delle Scienze, 7/A, 43100 Parma, Italy}

\begin{abstract}
We study the lattice N=1 Wess-Zumino model in two dimensions and we construct a
sequence $\rho^{(L)}$ of exact lower bounds on its ground state energy density $\rho$,
converging to $\rho$ in the limit $L\to\infty$. The bounds $\rho^{(L)}$ can be computed
numerically on a finite lattice with $L$ sites and can be exploited to discuss
dynamical symmetry breaking. The transition point is determined and compared
with recent results based on large-scale Green Function Monte Carlo simulations with 
good agreement.
\end{abstract}


\maketitle


\section{Introduction}

An important problem arising in the study of supersymmetric models is the
occurrence of non perturbative dynamical symmetry breaking~\cite{SUSYLattice}. 
The problem can be studied in $1+1$ dimensional lattice models where numerical tools 
are more effective and it should be easier to obtain a definite answer to the relevant 
questions.

In particular, the simplest theoretical laboratory is in our opinion 
the N=1 Wess-Zumino model that involves chiral superfields only and does not
present complications related to gauge invariance. Rigorous results in the continuum 
case can be found in~\cite{Jaffe}.
On the lattice, accurate numerical results are available~\cite{WZus,TopCharge}, but nevertheless
a clean determination of the supersymmetry breaking transition remains rather elusive.

The numerical simulations in Ref.~\cite{WZus} were performed using the Green Function Monte Carlo
(GFMC) algorithm and strong-coupling expansions. The physics of the model is fully determined by 
a single function of the scalar field, called in the following {\em prepotential} $V(\varphi)$.
All the results for the model with cubic prepotential indicated unbroken supersymmetry. 
Dynamical supersymmetry breaking in the model with quadratic prepotential 
$V=\lambda_2\varphi^2 + \lambda_0$ was studied performing
numerical simulations along a line of constant $\lambda_2$, and confirmed the existence of two
phases: a phase of broken SUSY with unbroken discrete $Z_2$ at high $\lambda_0$ and a phase of unbroken SUSY with
broken $Z_2$ at low $\lambda_0$, separated by a single phase transition.
We also studied the approach to the continuum limit in the model with quadratic
prepotential performing numerical simulations along a 1-loop renormalization group trajectory
in the phase of broken supersymmetry. 

The aim of the present paper is to propose a new approach to the study of the transition point.
The method is based on the calculation of rigorous lower bounds on the ground state 
energy density in the {\em infinite-lattice\/} limit. Such bounds are useful in the discussion of the 
supersymmetry phase diagram as follows from a rather simple argument that we sketch here and expand later in the
paper.
The lattice version of the Wess-Zumino model conserves
enough supersymmetry to prove that the ground state has a non negative energy density $\rho\ge 0$, as its continuum limit.
Moreover the ground state is supersymmetric if and only if $\rho=0$,
whereas it breaks (dynamically) supersymmetry if $\rho>0$. Therefore,
if an exact positive lower bound $\rho_{\rm LB}$ is found with 
$0< \rho_{\rm LB} \le \rho$, we can claim that supersymmetry is broken.

In the paper we find a sequence $\rho^{(L)}$ of exact lower bounds representing the ground state
energy densities of modified lattice Hamiltonians describing a cluster of $L$ sites,
which can be computed numerically. We stress that $\rho^{(L)}$ is an exact bound for each $L$ and 
moreover that it is consistent in the sense that $\rho^{(L)}\to \rho$ from below for $L\to\infty$.
This features will permit to draw conclusions on the critical values of the coupling constants in the 
infinite-volume limit.

The plan of the paper is the following: in Sec.~\ref{sec:model} we review the lattice formulation of the 
N=1 Wess-Zumino model. In Sec.~\ref{sec:lb} we derive the lower bound. In Sec.~\ref{sec:numerics} we review
the available results on the supersymmetric transition of the model and discuss our new 
results. Conclusions are summarized in Sec.~\ref{sec:conclusions}. Finally, in Appendix~\ref{app:free} we discuss the
bound in the free limit and in Appendix~\ref{app:worldline} we collect a few useful
technical details on the simulation algorithm used for the actual evaluation of the bound.

\section{The N=1 Wess-Zumino Model on the Lattice}
\label{sec:model}

The most general SUSY algebra in two dimensions has
$N$ left-handed fermionic generators $\{Q_L^A\}_{A=1,\dots, N}$ and
$\overline{N}$ right-handed fermionic generators $\{Q_R^A\}_{A=1,\dots,\overline N}$
and is denoted by $(N, \overline N)$. The bosonic generators are the
components of the two-momentum $(P^0, P^1)$ and central
charges $T^{AB}$. In the $(N,\overline N) = (1,1)$ case we denote
$Q_{1,2} \equiv Q_R^1\pm Q_L^1$ and find
$
\{Q_a, Q_b\} = 2(H\ {\bf 1} + P\sigma^1 + T\sigma^3)_{ab},
$
where $\sigma^i$ are the Pauli matrices, $(P^0, P^1)\equiv (H, P)$ 
and $T \equiv T^{11}$.
The Wess-Zumino model realizes the above algebra on a real chiral multiplet with a real scalar component 
$\varphi$ and a Majorana fermion with components $\psi_{1,2}$. The supercharges are
\be
Q_{1,2} = \int dx 
\left[p\psi_{1,2}
-\left(\frac{\partial\varphi}{\partial x}\pm V(\varphi)\right)\psi_{2,1}\right] ,
\ee
where $p(x)$ is the momentum operator conjugate to $\varphi(x)$
and $V(\varphi(x))$ is an arbitrary function called {\em prepotential} in the following.

On the lattice there are no continuous translation and we can only
preserve a SUSY subalgebra~\cite{Elitzur}.
We pick one of the supercharges, say $Q_1$, build a discretized version $Q_L$ and finally 
define the lattice Hamiltonian to be 
$H = Q_L^2$.

The explicit lattice model is built by considering a spatial
lattice with $L$ sites and open boundary conditions. On each site we place 
a real scalar field $\varphi_n$ together with its conjugate momentum $p_n$ such 
that $[p_n, \varphi_m] = -i\delta_{n,m}$.  The associated fermion is a Majorana fermion
$\psi_{a, n}$ with $a=1, 2$ and $\{\psi_{a, n}, \psi_{b, m}\} =
\delta_{a,b}\delta_{n,m}$ , $\psi_{a,n}^\dagger = \psi_{a,n}$. The
discretized supercharge is 
\be
Q_L = \sum_{n=1}^L\left [
p_n\psi_{1,n}-\left(\frac{\varphi_{n+1}-\varphi_{n-1}}{2}
+V(\varphi_n)\right)\psi_{2,n}\right] .
\ee
Following \cite{RanftSchiller} we replace the two Majorana fermion operators
with a single Dirac operator $c$ satisfying canonical
anticommutation rules, i.e.,
$\{c_n, c_m\} = 0$, $\{c_n,c_m^\dagger\} = \delta_{n,m}$:
\be
\psi_{\{1,2\},n} = \frac{(-1)^n \mp i}{2i^n}(c_n^\dagger\pm ic_n) .
\ee
The Hamiltonian is $H = H_B+H_F$ with 
\ba
H_B &=& \sum_{n=1}^L\left\{ \frac 1 2 p_n^2 + \frac 1
2\left(\frac{\varphi_{n+1}-\varphi_{n-1}}{2} + V(\varphi_n)\right)^2
\right\} , \\
H_F &=& \sum_{n=1}^L\left\{-\frac 1 2 (c^\dagger_n c_{n+1} + h.c.) + 
(-1)^n V'(\varphi_n) \left(c^\dagger_nc_n-\frac 1 2 \right)
\right\} . \nonumber
\ea
It conserves the total fermion number
\be
N_f = \sum_{n=1}^L N_n,\qquad N_n = c^\dagger_nc_n ,
\ee
and can be examined in each sector with definite $N_f$ separately. In this work, we shall
always consider the half-filled sector where the ground state is expected to lie. 
We remark that in this sector there exists a particle-hole symmetry which will play a role
in the following.

The relevant quantity for our analysis is the ground state energy density
$\rho$ evaluated on the infinite lattice 
\be
\label{rho}
\rho = \lim_{L\to\infty}\frac{E_0(L)}{L} .
\ee
It can be used to tell between the two phases of the model: supersymmetric with $\rho=0$ or 
broken with $\rho > 0$.
In the next Section we shall obtain  rigorous lower bounds on $\rho$ and exploit 
them to determine the phase.

\section{A Family of Lower Bounds on the energy density}
\label{sec:lb}

\subsection{Derivation of the bounds}

Given a translation-invariant Hamiltonian $H$ on the lattice $\mathbb{Z}$
it is possible to obtain lower bounds on its ground state energy density from 
a cluster decomposition of $H$. Briefly, given a suitable finite 
sublattice $\Lambda\subset\mathbb{Z}^d$, 
it is possible to introduce a modified Hamiltonian 
$\widetilde H$ restricted to $\Lambda$ such that its energy density $\rho_\Lambda$ bounds $\rho$ from 
below. The difference between $H$ and $\widetilde H$ is almost trivial and amounts to 
a simple rescaling of its coupling constants.
This simple fact was already noted in 1951 by Anderson~\cite{Anderson} and has been recently
improved and exploited in purely fermionic extended Hubbard models~\cite{BoundExtended}.
The derivation is quite general and applies to generic models with finite range interactions
as we now review. 

As the sublattice $\Lambda$ invades $\mathbb{Z}$, the bound $\rho_\Lambda$ is expected to improve
its accuracy with $\rho_\Lambda\to\rho$ in the infinite $\Lambda$ limit.
This property, that we call {\em consistency} in the following, will play a crucial 
role in the analysis of the Wess-Zumino model and, for this reason, we shall discuss it 
in some details.

As a preliminary discussion, it is expedient to make a general remark. Let $H$ be any
translation-invariant Hamiltonian on the $d$-dimensional lattice $\mathbb{Z}^d$, the only 
restriction on $H$ being that of a finite-range interaction between non overlapping subsets.
For instance, $H$ could include nearest-neighbor interaction terms,
next-to-nearest-neighbor couplings, and so on until some finite distance.

Let $\Lambda$ be any finite subset of the lattice, $H_\Lambda$ the restriction of $H$ to $\Lambda$
and $E_0(\Lambda)$ the energy of its ground state $|\psi_{0,\Lambda}\rangle$. We define
\be
\rho \stackrel{\rm def}{=} \lim_{\Lambda\to\infty}\frac{E_0(\Lambda)}{|\Lambda|} .
\ee
and show that 
\be
\label{infinitevolume}
\rho = \lim_{\Lambda\to\infty}\frac{\langle H_\Lambda\rangle}{|\Lambda|},
\ee
where for each local observable $\cal O$,
\be
\langle{\cal O}\rangle = \lim_{\Lambda\to\infty}\langle\psi_{0,\Lambda}|{\cal O}|\psi_{0,\Lambda}\rangle = 
\lim_{\Lambda\to\infty}\langle{\cal O}\rangle_\Lambda .
\ee
The expectation value $\langle{\cal O}\rangle$ means the average of
$\cal O$ in the infinite-volume ground state of $H$.
The proof of Eq.~(\ref{infinitevolume}) is easy. Since $\langle H_\Lambda\rangle = \lim_{\Lambda'}\langle H_\Lambda\rangle_{\Lambda'}
\ge E_0(\Lambda)$, we have the trivial inequality
\be
\lim_{\Lambda\to\infty}\frac{1}{|\Lambda|} \langle H_\Lambda \rangle \ge \lim_{\Lambda\to\infty}\frac{E_0(\Lambda)}{|\Lambda|}  = \rho .
\ee
We'll show now that the opposite inequality holds, and consequently that Eq.~(\ref{infinitevolume}) is true.

\begin{center}
\begin{figure}[htb]
\epsfig{width=16cm,angle=90,file=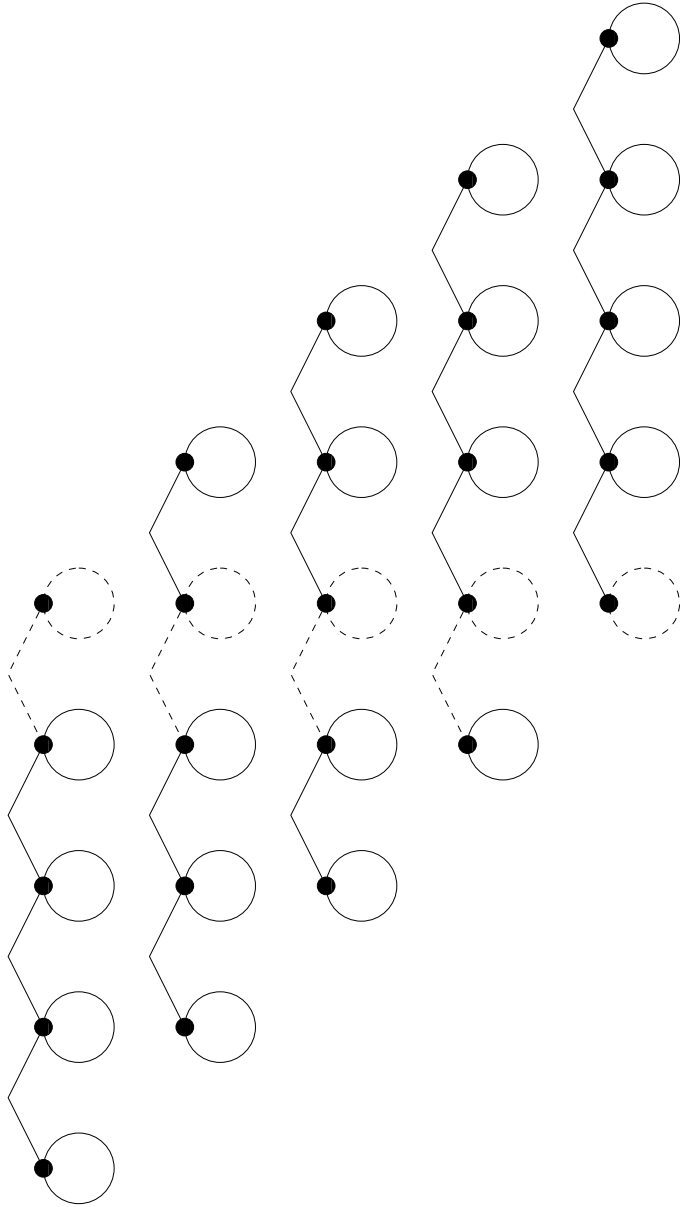}
\caption{Explanation of the factor $L/(L-1)$ in the nearest-neighbor terms of the modified
cluster Hamiltonian Eq.~(\ref{clusterham}). The black dots are the sites of a cluster with 5 sites. Circles and the arcs
denote on-site and nearest-neighbor interactions. If the cluster Hamiltonian is shifted over all positions of an infinite
lattice and all the contributions are summed, the dashed on-site interactions appear with multiplicity 5 and the dashed arc
with multiplicity 4. In the general case, the ratio is precisely $L/(L-1)$.}
\label{fig:counting}
\end{figure}
\end{center}

Due to the finite range of the interaction between non overlapping subsets of $\mathbb{Z}^d$, when $\Lambda'$ is a sufficiently
large region containing the bounded subset $\Lambda$ in its interior, we have
\be
H_{\Lambda'} = H_{\Lambda} + H_{\Lambda'\backslash \Lambda} + H_{\partial\Lambda},
\ee
where $\partial\Lambda$ is the boundary of $\Lambda$. Consequently, 
\be
\label{separation}
E_0(\Lambda') = E_0(\Lambda) + E_0(\Lambda'\backslash\Lambda) + {\cal O}(\partial\Lambda) .
\ee
Now, 
\be
E_0(\Lambda') = \langle H_{\Lambda'}\rangle_{\Lambda'} = \langle H_\Lambda + H_{\Lambda'\backslash\Lambda}
+H_{\partial\Lambda}\rangle_{\Lambda'} \ge \langle H_\Lambda\rangle_{\Lambda'} + E_0(\Lambda'\backslash\Lambda) + 
\langle H_{\partial\Lambda}\rangle_{\Lambda'} ,
\ee
and therefore, by exploiting Eq.~(\ref{separation}), we get the inequality
\be
\langle H_\Lambda\rangle \le E_0(\Lambda) + {\cal O}(\partial \Lambda) ,
\ee
when $\Lambda'\to\infty$. Therefore
\be
\lim_{\Lambda\to\infty}\frac{1}{|\Lambda|}\langle H_\Lambda\rangle \le \lim_{\Lambda\to\infty}\frac{E_0(\Lambda)}
{|\Lambda|} = \rho .
\ee
Coming back to the lower bound or the ground state energy density,
in order to illustrate the method we postpone the discussion of the Wess-Zumino model and 
consider first a generic lattice Hamiltonian with only on-site and nearest 
neighbor interactions 
\be
H = \sum_{n=-\infty}^\infty (H_n^{(0)} + H_n^{(1)}) ,
\ee
where $H_n^{(k)}$ involves operators at site $n$, $n+1$, \dots, $n+k$. 
We assume translation invariance, i.e., $H_n^{(k)}$ has an explicit form 
that does not depend on the base site $n$. 
We denote by $H_{s, N}$ the restriction of $H$ to a (connected) sublattice starting at site $s$ and ending at $s+N-1$, i.e., a cluster of $N$ sites
starting at $s$. We define it precisely as
\be
H_{s,N} = \sum_{n=s}^{s+N-1} H_n^{(0)} + \sum_{n=s}^{s+N-2} H_n^{(1)} .
\ee
To derive the bound it is convenient to define the following cluster Hamiltonian
where the relative weight of the on site and nearest neighbor terms has been modified
\be
\label{clusterham}
\widetilde H_{s, L} = \sum_{n=s}^{s+L-1} H_n^{(0)} + \frac{L}{L-1}\sum_{n=s}^{s+L-2} H_n^{(1)} .
\ee
We now superpose several copies of $\widetilde H_{s, L}$ in order to reconstruct the initial Hamiltonian. 
We begin with a fixed lattice of $N$ sites and define accordingly 
\be
\label{ClusterDecomposition}
\widetilde H^{(N,L)} = \sum_{n=1}^{N-L+1} \widetilde H_{n,L} .
\ee
We can expand the sum and rearrange terms obtaining
\ba
\label{Bulk}
\widetilde H^{(N,L)} &=& 
L\left(\sum_{n=L}^{N-L+1} H_n^{(0)} + \sum_{n=L}^{N-L} H_n^{(1)}\right) + {\cal B}, \\
&=& L H_{L, N-2L+2} + {\cal B}, \nonumber 
\ea
where $\cal B$ is a boundary term spanning a number of sites growing like $L$, but independent on $N$, 
located at the left and right ends of the lattice. 
The above results is made intuitive in Fig.~\ref{fig:counting} where we show the obvious origin of the factor $L/(L-1)$ in the 
nearest-neighbor terms of $\widetilde H_{\bullet, L}$.

It is now convenient to introduce 
\ba
E_0^{(N, L)} &=& \mbox{ground state energy of}\  \widetilde H^{(N,L)}, \\ 
\label{inter}
E_0^{(L)}    &=& \mbox{ground state energy of}\  \widetilde H_{\bullet, L}, \\
\rho^{(L)}   &=&  E_0^{(L)}/L .
\ea
where in Eq.~(\ref{inter}) we have not specified the initial site due to
translation invariance.  From Eq.~(\ref{ClusterDecomposition}) and
convexity we deduce the inequality
\be
E_0^{(N,L)} \ge (N-L+1) E_0^{(L)} ,
\ee
and dividing by $LN$ 
\be
\frac{E_0^{(N,L)}}{NL} \ge \frac{N-L+1}{N} \rho^{(L)} .
\ee
Let $\rho^*$ be the limit of the l.h.s.\ as $N\to\infty$ with fixed $L$. If we prove that 
$\rho \ge \rho^*$ we obtain the desired lower bound
\be
\label{bound}
\rho  \ge \rho^{(L)} .
\ee
To show this let us rewrite Eq.~(\ref{Bulk}) in the form 
\be
\frac{H_{L, N-2L+2}}{N} = \frac{\widetilde H^{(N,L)}}{LN}-\frac{\cal B}{LN} .
\ee
Taking the expectation value $\langle\cdot\rangle$ with respect to the ground state of the original Hamiltonian 
on the infinite lattice we find
\be
\label{tmp1}
\frac{\langle H_{L, N-2L+2}\rangle}{N} \ge \frac{E_0^{(N,L)}}{LN}-\frac{\langle \cal B\rangle}{LN} .
\ee
In the limit $N\to\infty$, the contribution of the
boundary terms vanishes because it does not depend on $N$ due to 
the translation invariance of  $\langle\cdot\rangle$. The l.h.s.\ of Eq.~(\ref{tmp1}) is
 $\rho$ in the limit, by the general remark we made at the beginning 
and we thus find $\rho \ge \rho^*$ completing the proof of Eq.~(\ref{bound}) since
$\rho \ge \rho^* \ge \rho^{(L)}$.

The bound is also consistent in the sense that 
\be
\label{consistency}
\lim_{L\to\infty} \rho^{(L)} = \rho ,
\ee
as it can be shown by an argument similar to the above one. To this aim, we write 
\be
\label{inverse}
\frac{1}{L} \widetilde H_{s,L} = \frac{1}{L} H_{s,L} + \frac{1}{L(L-1)}\sum_{n=s}^{s+L-2} H_n^{(1)} ,
\ee
and we take now the expectation value with respect to the vacuum of $\widetilde H_{s,L}$ 
in the infinite-lattice
limit $L\to\infty$. Since
\be
\lim_{L\to\infty}\rho^{(L)} = \lim_{L\to\infty}\frac{1}{L}\langle \widetilde H_{0,L}\rangle,
\ee
by the general remark we made at the beginning, we obtain from Eq.~(\ref{inverse}) the inequality
\be
\lim_{L\to\infty} \rho^{(L)} \ge \rho,
\ee
that combined with $\rho^{(L)} \le \rho$ gives Eq.~(\ref{consistency}). 

In the case of the Wess-Zumino model it is necessary to include also next-to-nearest couplings, but
the procedure is identical and we just sketch its main features. We write
\be
H = \sum_{n=-\infty}^\infty (H_n^{(0)} + H_n^{(1)} + H_n^{(2)}) ,
\ee
where $H_n^{(k)}$ involves operators at site $n$, $n+1$, \dots, $n+k$. We have explicitly
\ba
\label{sign}
H_n^{(0)} &=& \frac 1 2 p_n^2 + \frac 1 4 \varphi_n^2 + \frac 1 2 V(\varphi_n)^2 +(-1)^n V'(\varphi_n) 
\left(c^\dagger_nc_n-\frac 1 2 \right) ,\\
H_n^{(1)} &=& \frac 1 2 V(\varphi_n) \varphi_{n+1} - \frac 1 2 \varphi_n V(\varphi_{n+1})
-\frac 1 2 (c^\dagger_n c_{n+1} + h.c.) ,\\
H_n^{(2)} &=& -\frac 1 4 \varphi_n\varphi_{n+2} .
\ea
Again, we define
\be
\label{ClusterWessZumino}
\widetilde H_{s, L} =  \sum_{n=s}^{s+L-1} H_n^{(0)} + \frac{L}{L-1} \sum_{n=s}^{s+L-2} H_n^{(1)}
 + \frac{L}{L-2} \sum_{n=s}^{s+L-3} H_n^{(2)} .
\ee
Due to the presence of the term $(-1)^n$ in the fermion-boson coupling in Eq.~(\ref{sign}) the
Hamiltonians $\widetilde H_{s, L}$ for even and odd $s$ are different, but they are related by the 
particle-hole symmetry, as we remarked before. Therefore they have the same ground state energy
and the previous discussion applies. Still, there is a translation invariance on our lattice,
but only under shifts by two lattice spacings. In our staggered fermion formulation, it is the remnant 
of the continuum translation invariance.
As before, we prove for 
\be
\rho^{(L)} = \mbox{ground state energy of}\  \frac{1}{L} \widetilde H_{\bullet, L} , 
\ee
the relations
\be
\rho \ge \rho^{(L)},\quad {\rm with}\quad \lim_{L\to\infty} \rho^{(L)} = \rho.
\ee

\subsection{Relevance to the problem of supersymmetry breaking}

We now explain the way we exploit the sequence of bounds $\rho^{(L)}$ (indexed by the cluster size $L$)
to determine the phase at a particular point in the coupling constant space. To this aim, we compute numerically 
$\rho^{(L)}$ at various values of the cluster size $L$. If we find $\rho^{(L)}>0$ for some $L$,
we can immediately conclude that we are in the broken phase. If, on the other hand, we find a negative 
bound we cannot conclude in which phase we are. However, we know that $\rho^{(L)}\to \rho$ for $L\to\infty$
and the study of $\rho^{(L)}$ as a function of {\em both\/} $L$ and the coupling constants 
permit the identification of the phase in all cases.
We shall discuss this strategy in 
more details and for the particular case of the Wess-Zumino model in the next Section dedicated to 
the presentation of our numerical results.

We remark that the calculation of $\rho^{(L)}$ is numerically feasible because it requires to 
determine the ground state energy of a Hamiltonian quite similar to $H$ and defined on a finite lattice 
with $L$ sites. In particular, it is not necessary to deal with the difficult problem of 
computing any infinite-size limit. In other words, we are able to transfer the information obtained 
in finite volume to a result holding on the infinite lattice. As a simple semi-analytical illustration
of the technique, we report in Appendix~\ref{app:free} the evaluation of the bound in the
free model with zero prepotential.

\section{Numerical Results}
\label{sec:numerics}

\begin{center}
\begin{figure}[htb]
\epsfig{width=16cm,angle=0,file=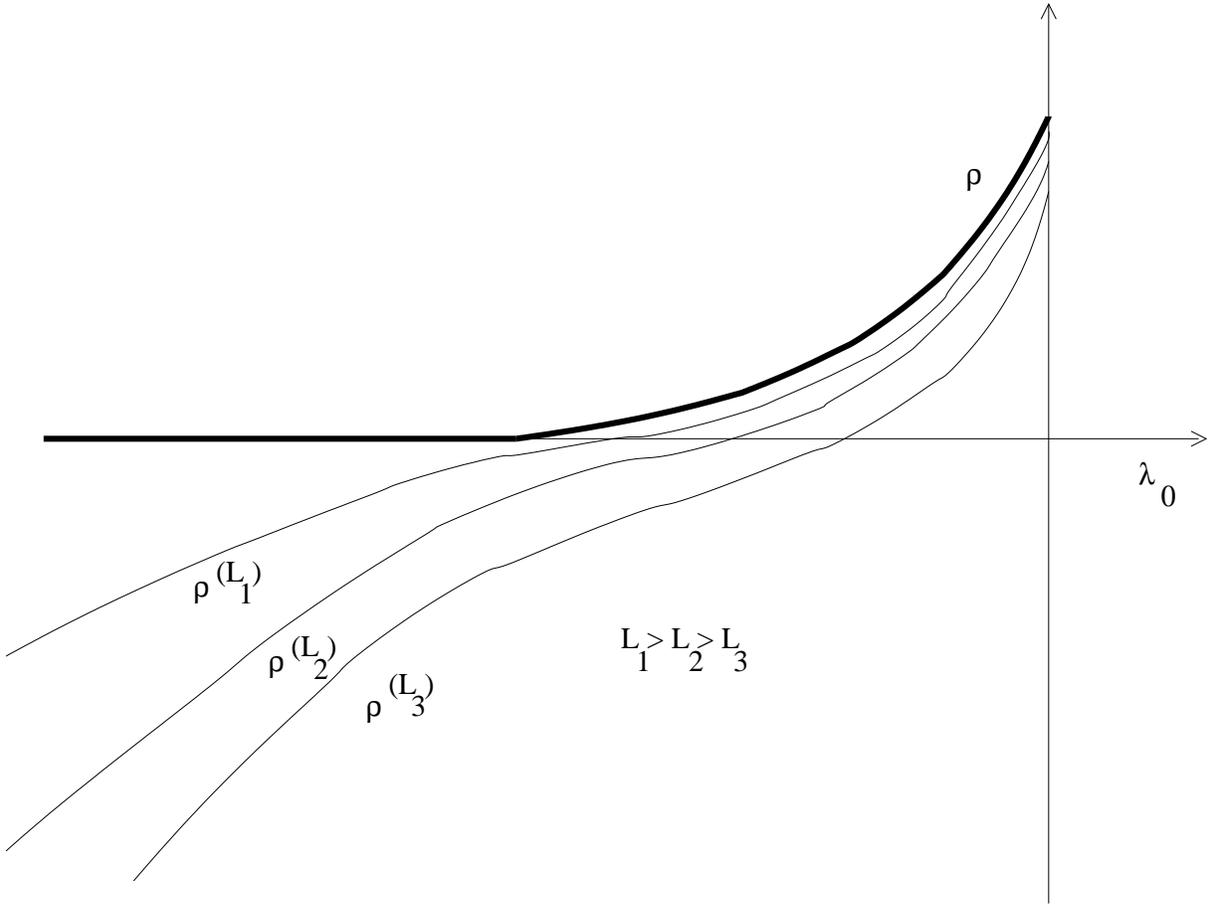}
\caption{Qualitative plot of the functions $\rho(\lambda_0)$ and $\rho^{(L)}(\lambda_0)$.}
\label{fig:qualitative}
\end{figure}
\end{center}

To test the effectiveness of the proposed bound and its relevance to the problem of locating the supersymmetric
transition in the Wess-Zumino model, we study in details the case of a quadratic prepotential
\be
\label{quadratic}
V(\varphi) = \lambda_2 \varphi^2 + \lambda_0 , 
\ee
and discuss the dependence of $\rho$ on $\lambda_0$ at fixed $\lambda_2$. Indeed, in the continuum,
for a given $\lambda_2$, a general argument by Witten~\cite{Witten} and 
rigorous results in constructive quantum field theory~\cite{Jaffe} suggest the existence of a 
negative number $\lambda_0^*$ such that $\rho(\lambda_0)$ is positive when $\lambda_0 > \lambda_0^*$ and
it vanishes for $\lambda_0 < \lambda_0^*$. Such number $\lambda_0^*$ is the value  of $\lambda_0$
at which dynamical supersymmetry breaking occurs. We draw in Fig.~\ref{fig:qualitative} a reasonable qualitative 
pattern of the curves representing $\rho^{(L)}(\lambda_0)$. We see that a single zero is expected in $\rho^{(L)}(\lambda_0)$
at some $\lambda_0 = \lambda_0(L)$. Since $\lim_{L\to\infty}\rho^{(L)} = \rho$, we expect that 
$\lambda_0(L)\to\lambda_0^*$ for $L\to \infty$
allowing for a determination of the critical coupling $\lambda_0^*$.

We stress that the continuum limit of the model is obtained by following a Renormalization Group
trajectory that, in particular, requires the limit $\lambda_2\to 0$. This step has been partially
accomplished in Ref.~\cite{WZus} as explained in the next paragraph, but will not be pursued in the present 
paper. Here, we consider the lattice Wess-Zumino model at intermediate couplings and, we stress
again, in the infinite-lattice limit.

\subsection{Review of GFMC results}

Here we give a brief review of our results in Ref.~\cite{WZus}, in order to compare them 
with the results in the present paper. 
In Ref.~\cite{WZus} the analysis of supersymmetry breaking in a class of two dimensional lattice Wess-Zumino model (with open boundary conditions)
was performed using the GFMC algorithm. 
This method computes a numerical representation of the ground state on a finite lattice with $L$ sites 
in terms of the states carried by an ensemble of $K$ walkers. The need
to extrapolate to infinite $L$ and $K$ is the main source of systematic error.
As an example of odd prepotential the case $V=\varphi^3$ 
was investigated, measuring the ground-state energy and the supersymmetric Ward identity. 
Both of them gave a very convincing evidence for unbroken SUSY. 
In Ref.~\cite{WZus}, in order to study the supersymmetry breaking pattern, the more interesting case of even 
prepotential (\ref{quadratic}) was investigated.
In this case the model enjoys an approximate $Z_2$ symmetry which corresponds to the symmetry 
under the transformations $\varphi_n\to-\varphi_n$, $\chi_n\leftrightarrow\chi_n^\dagger$.
The theoretical expectation is that, for a fixed value of $\lambda_2$,
at high $\lambda_0$ the model is in a phase of broken SUSY and
unbroken $Z_2$; at low $\lambda_0$ it is in a phase of unbroken SUSY
and broken $Z_2$. The case $\lambda_2=0.5$ was investigated in detail.

The usual technique for the study of a phase transition is the
crossing method applied to the Binder cumulant, $B$, with a sensible
choice of magnetization (which is not completely trivial, since our
model is neither ferromagnetic nor antiferromagnetic and it doesn't
enjoy translation symmetry).
The crossing method consists in plotting $B$ vs.\ $\lambda_0$ for
several values of $L$. The crossing point $\lambda_0^{\rm cr}(L_1,L_2)$, determined by the condition
\[
B(\lambda_0^{\rm cr},L_1) = B(\lambda_0^{\rm cr},L_2)
\]
is an estimator of $\lambda_0^*$.
The convergence is dominated by the critical exponent $\nu$ of the correlation length and by the 
critical exponent $\omega$ of the leading corrections to scaling 
\[
\lambda_0^{\rm cr}(L_1,L_2) = \lambda_0^{(c)} + 
O(L_1^{-\omega-1/\nu},L_2^{-\omega-1/\nu});
\]
we expect the phase transition we are studying to be in the Ising universality class, for
which $\nu=1$ and $\omega=2$, and therefore we
expect fast convergence $\lambda_0^{\rm cr}\to \lambda_0^*$.
The results \cite{WZus} indicate $\lambda_{0, \rm GFMC}^* = -0.48\pm0.01$.

It is possible to study the phase transition by looking at the connected correlation function 
$G_d=\langle\varphi_n\varphi_{m}\rangle_c$, averaged over all $n,m$
pairs with $|m-n|=d$, excluding pairs for which $m$ or $n$ is close
to the border.  Even  
and odd $d$ may correspond to different physical channels.
In Ref.~\cite{WZus} $G_d$ is fitted to the form $\exp[-a_1 - a_2 d + a_3/(d+10)]$,
separately for even and odd $d$. 
In the broken phase, we have small but nonzero $a_2$, and we observe equivalence of the even-
and odd-$d$ channels, while in the unbroken phase, $a_2$ is larger, and the even- and odd-$d$ 
channels are different \cite{WZus}. 
The difference between the two phases is apparent, e.g., in the plot
of $a_2$ vs.\ $\lambda_0$. The data presented in Ref.~\cite{WZus} confirms the quoted value of 
$\lambda_0^*$
\footnote{
Other methods to estimate $\lambda_0^*$ were discussed in Ref.~\cite{WZus}, but with quite larger 
errors. In particular a study of the bosonic fields effective potential 
suggested $\lambda_0^*\simeq-0.40$, whereas an extrapolation of $E_0/L$
to infinite $K$ and $L$ give $\lambda_0^*\sim-0.53$, with a rather large uncertainty.
}.

\subsection{Results for the lower bound $\rho^{(L)}(\lambda_0)$}

We consider the quadratic prepotential Eq.~(\ref{quadratic}) at the fixed value $\lambda_2 = 0.5$.
As discussed above, the properties of the bound $\rho^{(L)}(\lambda_0)$ guarantee that for large enough 
$L$ it must have a single zero $\lambda_0^*(L)$ converging to $\lambda_0$ as $L\to\infty$. In any case
for each $L$ we can claim that $\lambda_0^* > \lambda_0^*(L)$. To obtain a numerical estimate of 
$\rho^{(L)}(\lambda_0)$ we exploit the so-called worldline path integral (WLPI) algorithm. We discuss it in some details
in Appendix~\ref{app:worldline} and we just recall here its basic features in order to introduce the 
systematic errors of the analysis. The WLPI algorithm computes numerically the quantity
\be
\rho^{(L)}(\beta, T) = \frac{1}{L}\frac{\mbox{Tr}\{ H\ (e^{-\frac{\beta}{T} H_1}e^{-\frac{\beta}{T} H_2})^T\}}
{\mbox{Tr}\{ (e^{-\frac{\beta}{T} H_1}e^{-\frac{\beta}{T} H_2})^T\}} ,
\ee
where the Hamiltonian for a cluster of $L$ sites (\ref{ClusterWessZumino}) is written $H=H_1+H_2$ 
by separating in a convenient way the various bosonic and fermionic operators in the subhamiltonians $H_1$ and $H_2$.
The desired lower bound is obtained by the double extrapolation
\be
\rho^{(L)} = \lim_{\beta\to\infty}\lim_{T\to\infty} \rho^{(L)}(\beta, T),
\ee
with polynomial convergence $\sim 1/T$ in $T$ and exponential in $\beta$. Numerically, we determined 
$\rho^{(L)}(\beta, T)$ for various values of $\beta$ and $T$ and a set of $\lambda_0$ that should include
the transition point, at least according to the GFMC results. 

\begin{center}
\begin{figure}[htb]
\epsfig{width=20cm,angle=90,file=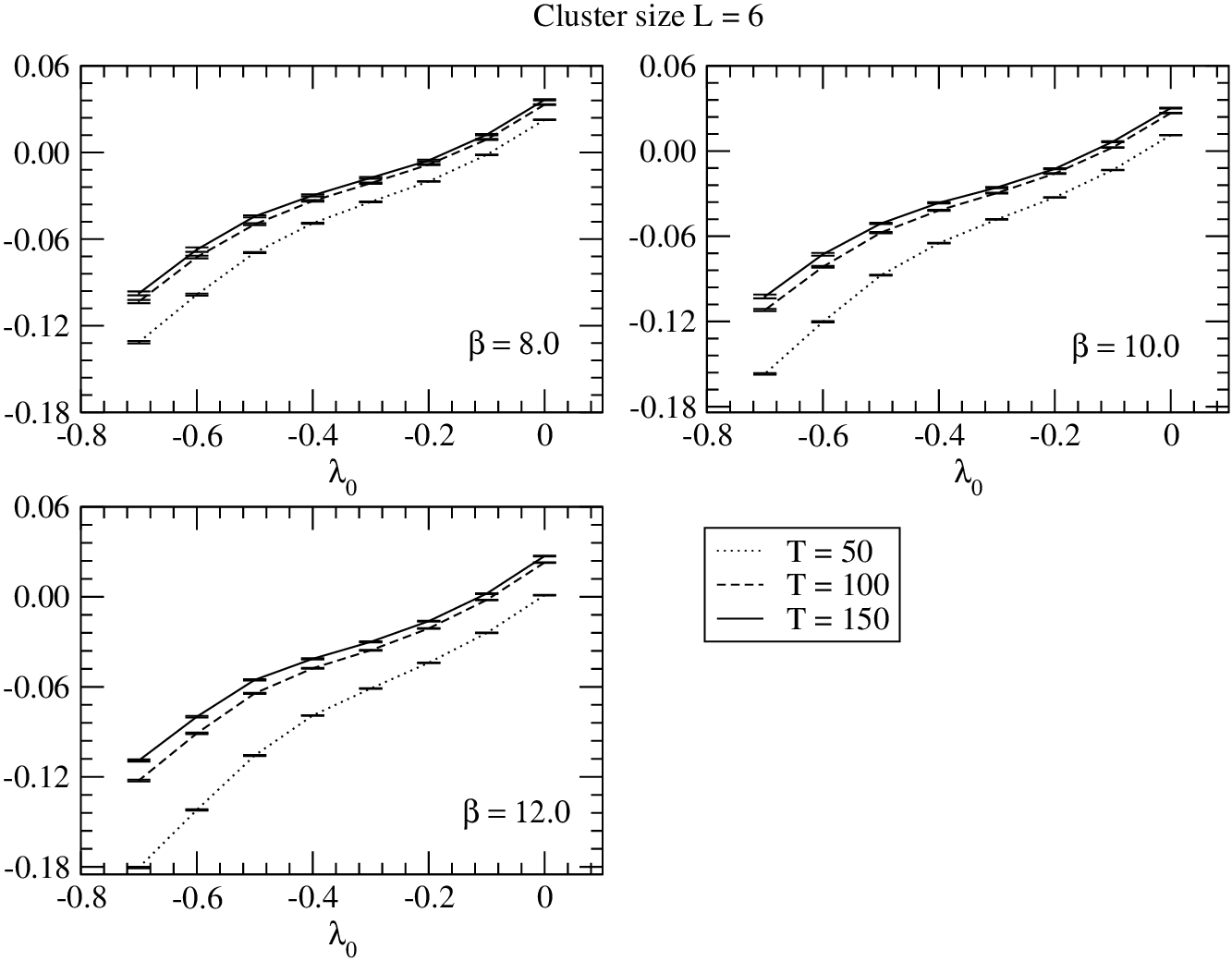}
\caption{Plot of the energy lower bound $\rho^{(L)}(\beta, T)$ at $L=6$.}
\label{fig:L6:raw}
\end{figure}
\end{center}

\begin{center}
\begin{figure}[htb]
\epsfig{width=20cm,angle=90,file=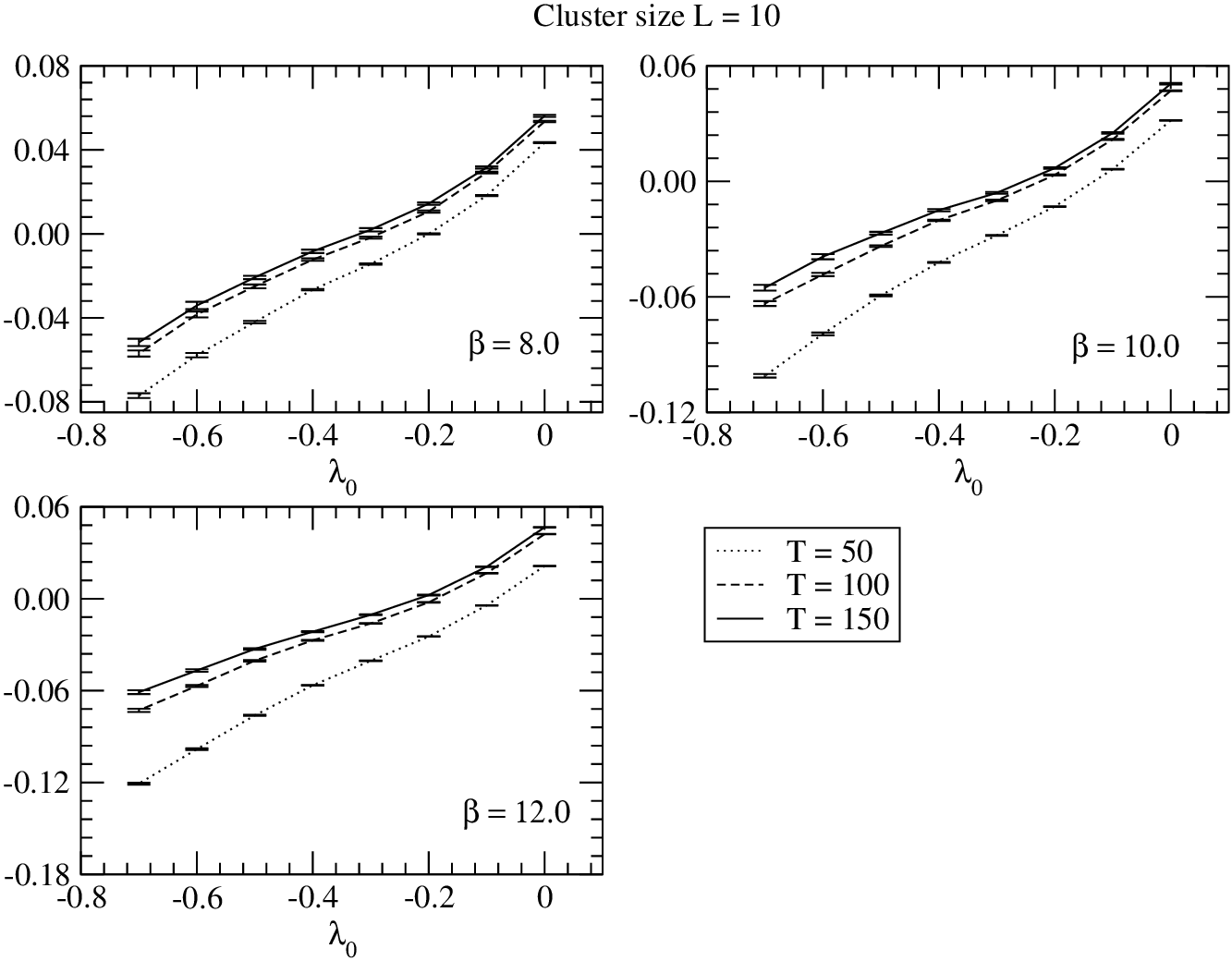}
\caption{Plot of the energy lower bound $\rho^{(L)}(\beta, T)$ at $L=10$.}
\label{fig:L10:raw}
\end{figure}
\end{center}

\begin{center}
\begin{figure}[htb]
\epsfig{width=20cm,angle=90,file=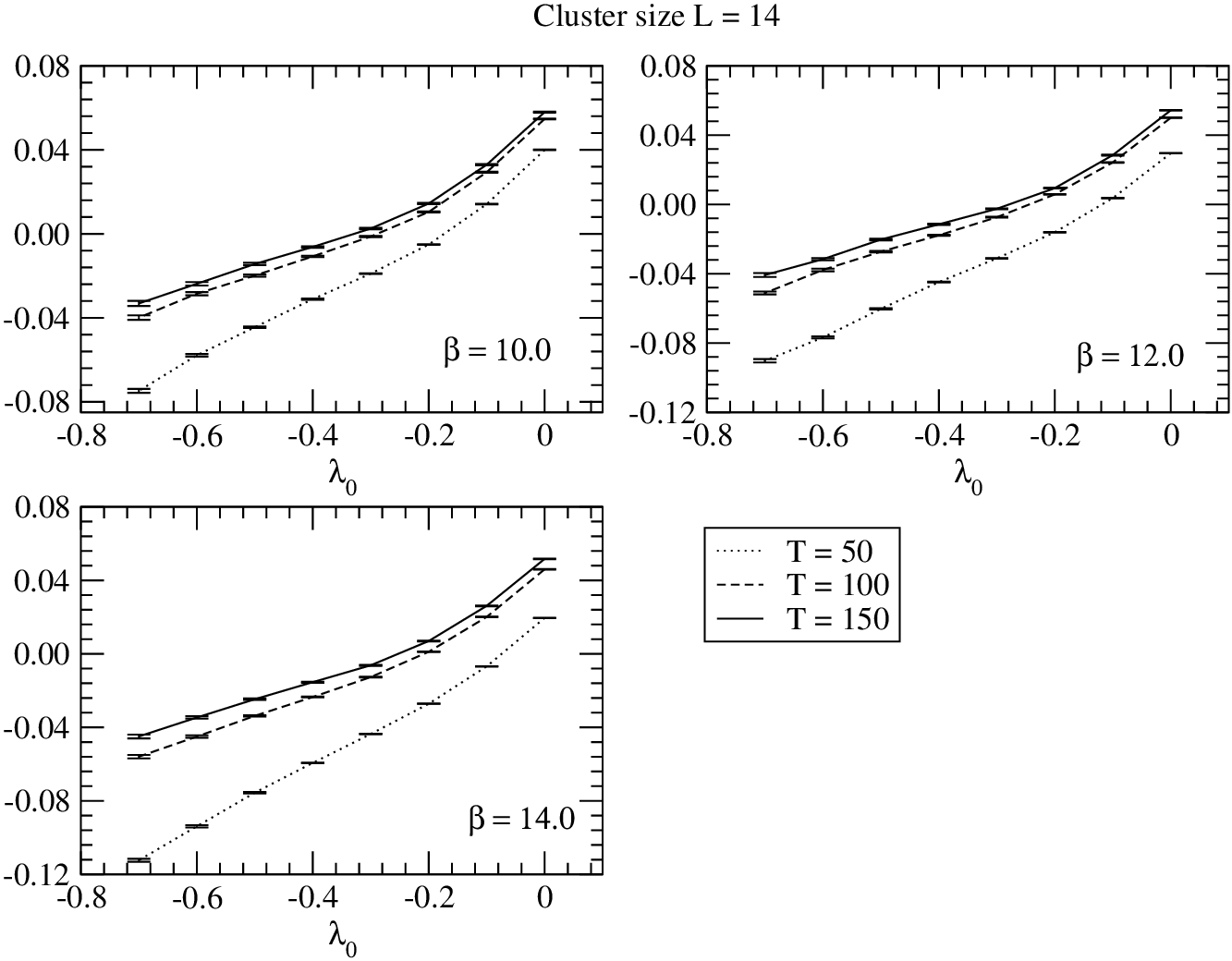}
\caption{Plot of the energy lower bound $\rho^{(L)}(\beta, T)$ at $L=14$.}
\label{fig:L14:raw}
\end{figure}
\end{center}

\begin{center}
\begin{figure}[htb]
\epsfig{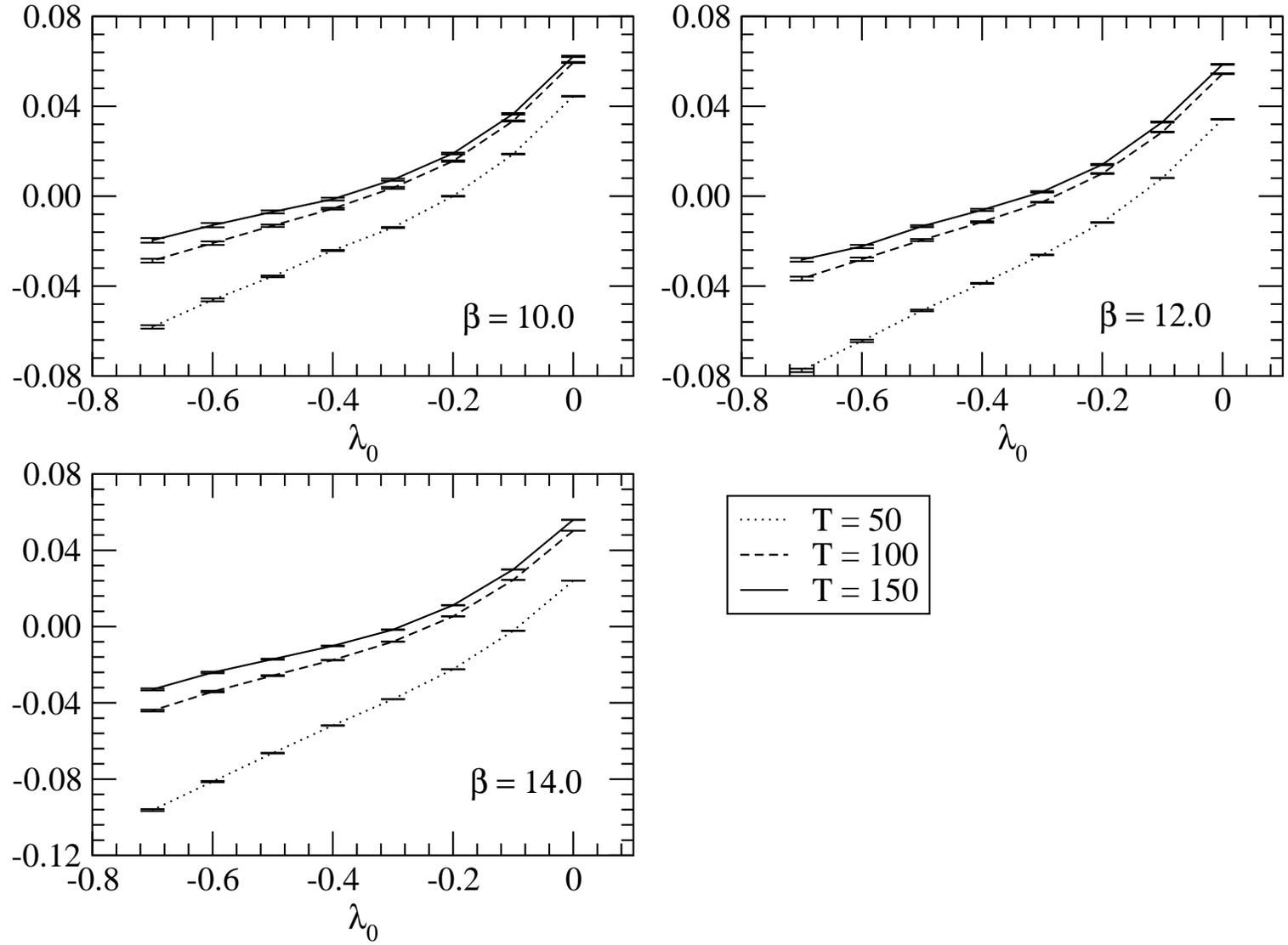}
\caption{Plot of the energy lower bound $\rho^{(L)}(\beta, T)$ at $L=18$.}
\label{fig:L18:raw}
\end{figure}
\end{center}

\begin{center}
\begin{figure}[htb]
\epsfig{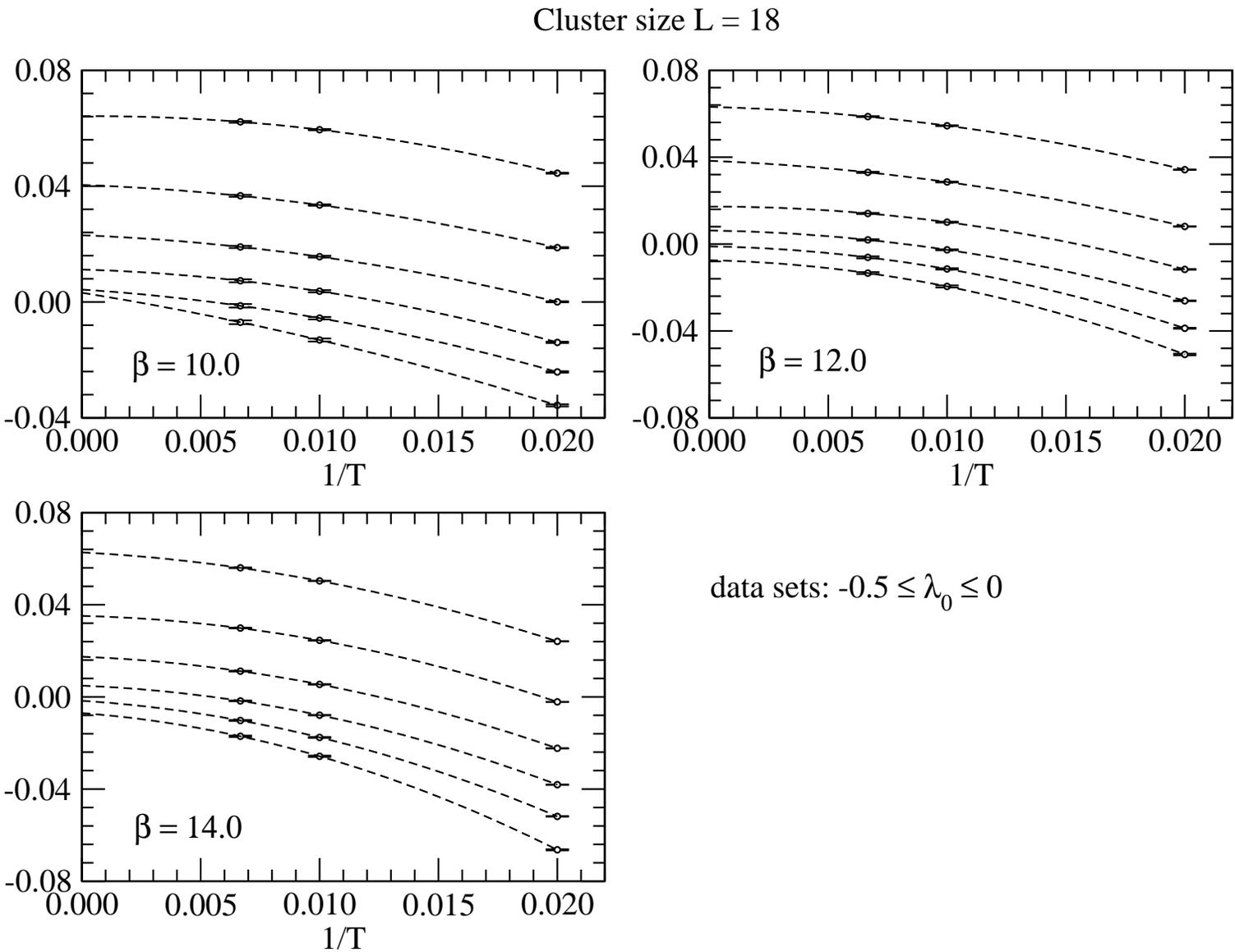}
\caption{An example of the extrapolation of $\rho^{(L)}(\beta, T)$ to the
$T\to\infty$ limit. The curves are quadratic in $1/T$ and for each $\beta$ they 
correspond to $-0.5 \le \lambda_0\le 0$ in uniform steps $\Delta\lambda_0 = 0.1$.}
\label{fig:L18:fit}
\end{figure}
\end{center}

\begin{center}
\begin{figure}[htb]
\epsfig{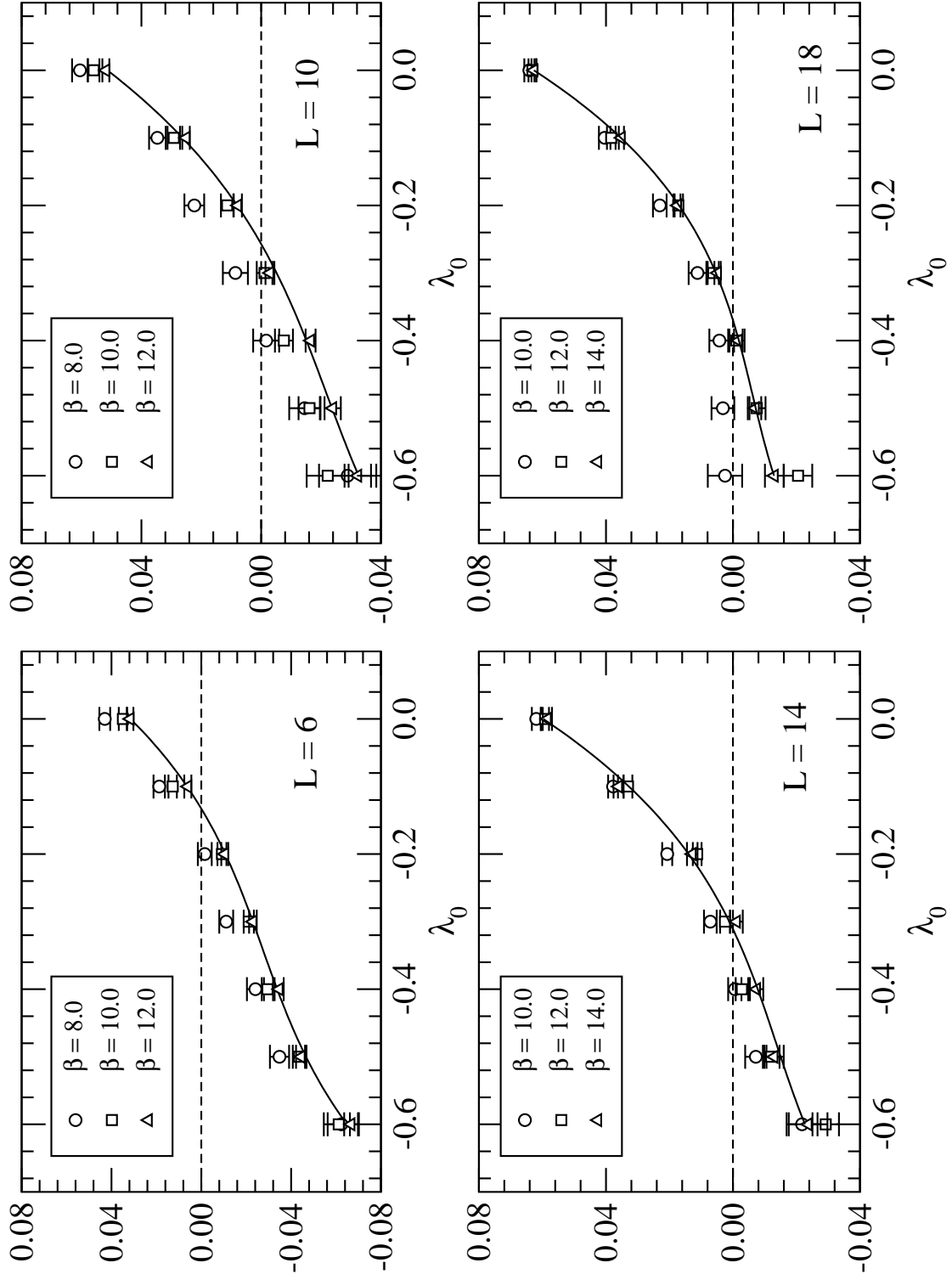}
\caption{Extrapolated bound $\rho^{(L)}(\beta) = \lim_{T\to\infty} \rho^{(L)}(\beta, T)$. As one can see, 
for each cluster size $L$, the two curves at the highest values of $\beta$ almost coincide. 
We assume that the curve with the highest $\beta$ describes correctly the zero temperature limit, i.e., the
quantum expectation value over the ground state. The solid line is a cubic spline through data.}
\label{fig:extrapolation1}
\end{figure}
\end{center}

\begin{center}
\begin{figure}[htb]
\epsfig{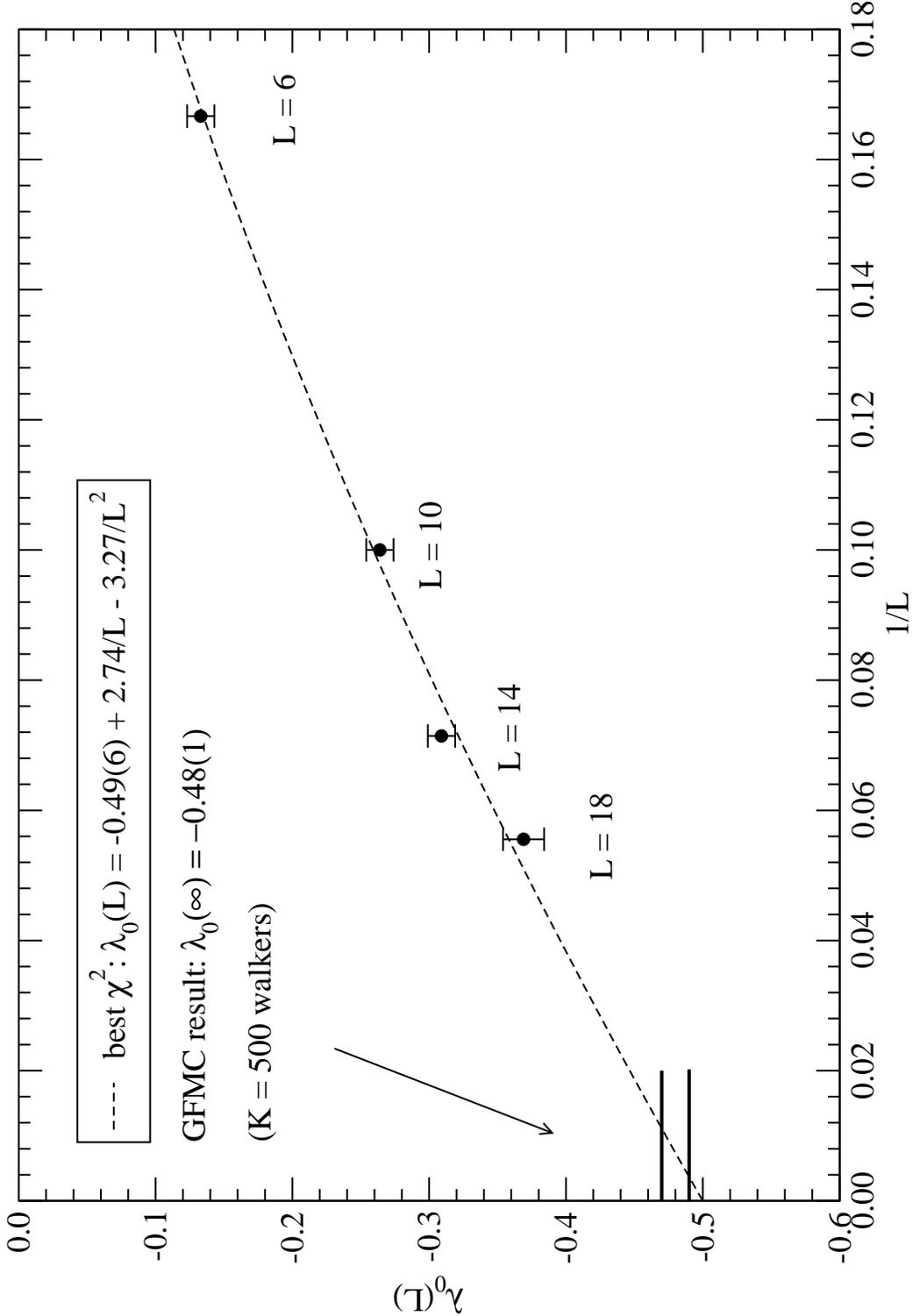}
\caption{The results shown in the previous Figure allow to define an upper bound $\lambda_0^*(L)$ on the
critical coupling. Consistency of this bound predicts that  $\lambda_0^*(L)$ converges as $L\to \infty$ to the 
infinite-lattice critical coupling. In the Figure, we show a best fit with a quadratic polynomial in $1/L$
together with the best GFMC result obtained with $K=500$ walkers (see Ref.~\cite{WZus} for details on GFMC).}
\label{fig:extrapolation2}
\end{figure}
\end{center}

Our basic numerical results are shown in Figs.~\ref{fig:L6:raw}--\ref{fig:L18:raw}, where we plot the function 
$\rho^{(L)}(\beta, T)$ for $L = 6$--$18$, various $\beta$ and $T=50$, $100$, $150$. In order to avoid fermionic sign 
problems we need $L=4k+2$. In this case, at half filling, when a fermion hops through the lattice it passes over
$L/2-1 = 2k$ fermions, i.e., an even number giving no signs in the partition function. The extrapolation to $T\to \infty$
is quadratic in $1/T$ and a sample case is shown in Fig.~\ref{fig:L18:fit}. The results are shown in Fig.~\ref{fig:extrapolation1}.
The chosen values of $\beta$ are such that the curves corresponding to the highest $\beta$ values can be 
taken as representatives of the $\beta\to\infty$ limit. This is rather safe due to the exponential convergence in $\beta$.
For all cluster sizes, we see that the energy lower bound behaves as expected: it is positive around $\lambda_0 = 0$ and 
decreases as $\lambda_0$ moves to the left. At a certain unique point $\lambda_0^*(L)$, the bound vanishes and 
remains negative for $\lambda_0 < \lambda_0^*(L)$. This means that supersymmetry breaking can be excluded 
for $\lambda_0 > \min_L\lambda_0^*(L)$. Also, consistency of the bound means that $\lambda_0^*(L)$ must converge
to the infinite-volume critical point as $L\to\infty$. Since the difference between the exact Hamiltonian and the 
one used to derive the bound is ${\cal O}(1/L)$, we can fit $\lambda_0^*(L)$ with a polynomial in $1/L$. This is 
shown in Fig.~\ref{fig:extrapolation2} where we also show the GFMC result. The best fit with a parabolic function 
gives $\lambda_0^* = -0.49\pm 0.06$ quite in agreement with the previous $\lambda_{0, \rm GFMC}^* = -0.48\pm0.01$.

The somewhat large error could of course 
be reduced by improving the $\beta$ and $T$ extrapolations and with additional statistics; however, we do not pursue 
further the numerical analysis, since the main aim of the present paper has
been to show the validity of the new proposal for the identification
of the infinite-lattice critical point, by a rather non-standard
method.

\begin{center}
\begin{figure}[htb]
\epsfig{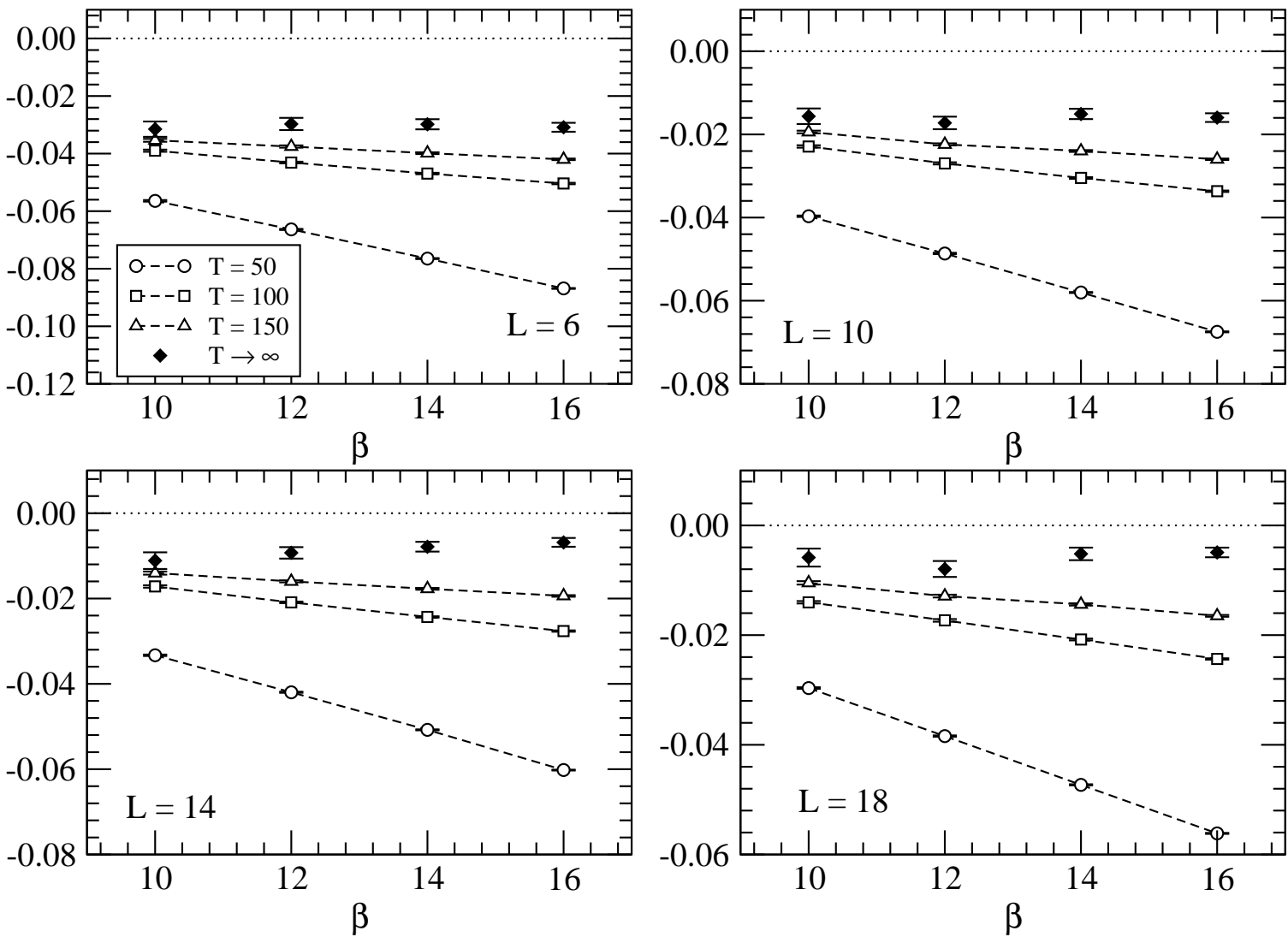}
\caption{Plot of the energy lower bound $\rho^{(L)}(\beta, T)$ at various $L$, $\beta$ and $T$ for
the cubic prepotential.}
\label{fig:cubicraw}
\end{figure}
\end{center}

As a concluding remark, we briefly discuss the cubic prepotential
$V(\varphi)~=~\frac 1 4 \varphi^3$, where general arguments
predict no supersymmetry breaking; this
has been confirmed by the GFMC analysis~\cite{WZus}. 
In Fig.~\ref{fig:cubicraw} we show the lower bound computed on the clusters
with $L=6$, $10$, $14$ and $18$ sites with $\beta = 10$, $12$, $14$ and three values $T=50$, $100$, $150$. 
We also show the results of a quadratic extrapolation in $1/T$ at fixed $\beta$. At a rather qualitative level
we remark two features of the Figure. First, the bound is definitely negative for all $L$; this is consistent
with our expectations because the infinite-volume energy density vanishes in this case. Second, the values at
large $\beta$ approach zero as $L$ increases as follows from the discussed consistency properties of the bound.

\section{Conclusions}
\label{sec:conclusions}

In the present paper we have studied the phase diagram of the lattice N=1 Wess-Zumino model with a particular
scalar potential for which we expect dynamical supersymmetry breaking to occur.
As it is known, the model can be put on the lattice with a certain amount of 
supersymmetry left unbroken by the space-time discretization. As an important consequence, 
the lattice model shares with its continuum counterpart the important property of having a 
non negative energy density $\rho\ge 0$. Also the vanishing of $\rho$ is a necessary and sufficient 
signal for unbroken supersymmetry, whereas $\rho > 0$ would imply a dynamically broken phase, that is expected
to exist on the infinite lattice.
All these facts trigger the interest toward a tight lower bound on $\rho$ holding, we repeat, 
on the infinite lattice.
We have proposed precisely a family of such bounds $\rho^{(L)}$ that can be computed in terms of 
the energy density of a suitable (non supersymmetric) local model defined on a 
cluster of $L$ sites. The family is consistent, i.e., $\rho^{(L)} \to \rho$ as $L\to\infty$. 

The analysis of $\rho^{(L)}$ as function of the cluster size $L$ and the coupling constants
permits to derive bounds on the critical values of the couplings where dynamical supersymmetry breaking
occurs. Also, extrapolation in the cluster size allows in principle to locate the transition in a rather 
straightforward way. 

To confirm the feasibility of the procedure we have computed numerically $\rho^{(L)}$,
with good agreement with existing previous studies based on extensive simulations
with the Green Function Monte Carlo method.
We thus conclude that the proposed approach is useful as an independent method to locate the supersymmetric 
transition.

As a final comment, we would like to emphasize what seems to us a general interesting point. The condition 
$\rho\ge 0$ in (globally) supersymmetric models is obviously and directly related to the issue of
symmetry breaking. Usually, this simple fact is not exploited in non-perturbative approaches based on 
standard simulations in the Lagrangian framework where the focus is mainly on correlation functions
to check for instance Ward identities. This is definitely not true in the (lattice) Hamiltonian formalism
where the ground state energy is typically the simplest observable that can be considered, at least in finite
volume, and the vanishing of $\rho$ is the most natural monitor of supersymmetry. As we have shown, 
analytical information on $\rho$, like rigorous bounds, can be immediately translated into a valid
tool to make the numerical investigation most effective.

\acknowledgments

Financial support from INFN, IS RM42 and PI12 is acknowledged.

\appendix

\section{Evaluation of the bound in the free model}
\label{app:free}

\begin{center}
\begin{figure}[htb]
\epsfig{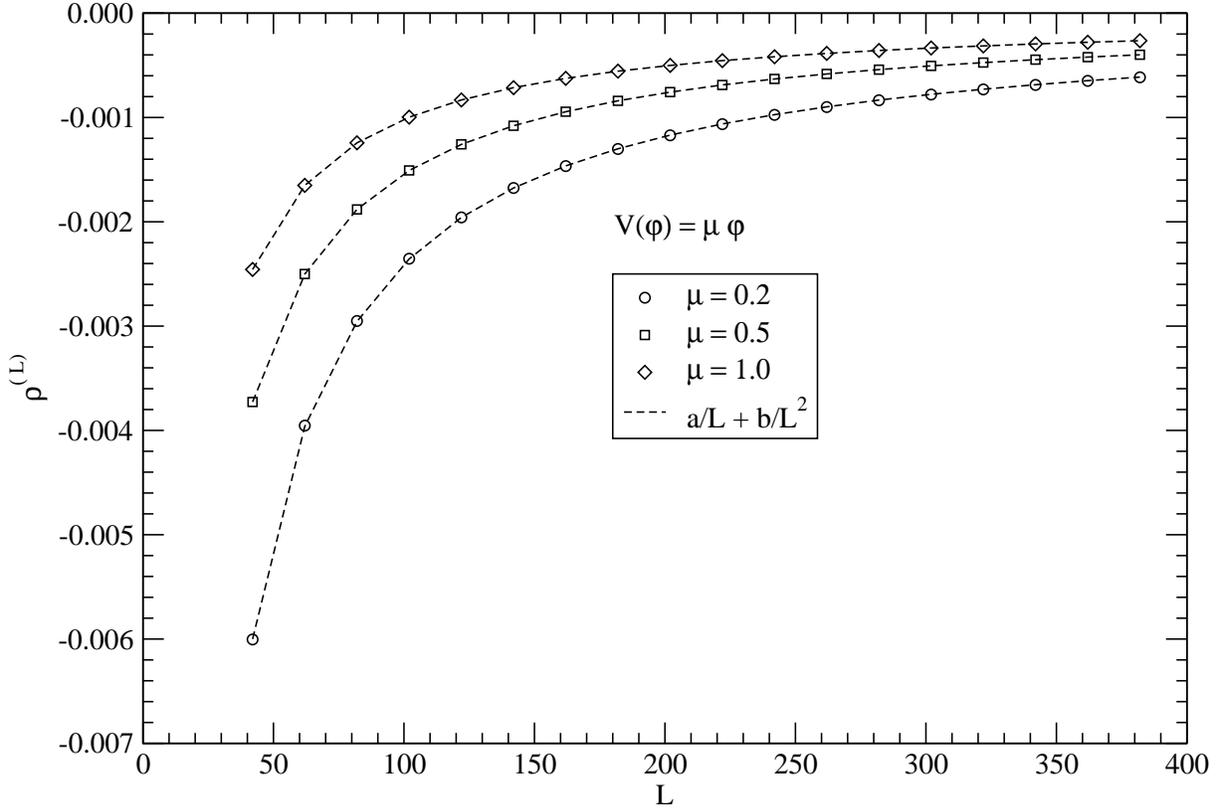}
\caption{Evaluation of the bound in the free limit $V(\varphi) = \mu\varphi$.}
\label{fig:free}
\end{figure}
\end{center}

The Wess-Zumino model with linear prepotential $V(\varphi) = \mu\varphi$ is free
with bosons decoupled from fermions. The Hamiltonian $\widetilde H_{1,L}$ can be written
\be
\widetilde H_{1,L} = \frac 1 2 \sum_{i=1}^L p_i^2 +\frac 1 2 \sum_{i,j=1}^L M^{(B)}_{ij}\varphi_i\varphi_j + 
\sum_{i,j=1}^L M^{(F)}_{ij} c^\dagger_i c_j ,
\ee
where the matrices $M^{(B)}$ and $M^{(F)}$ have elements
\be
M^{(B)}_{ij}=\left\{\begin{array}{cl}
-\frac{1}{4}\frac{L}{L-2} & |i-j|=2 \\
\frac{1}{2}+\mu^2         & i=j \\
0                         & \mbox{else}
\end{array}\right. , \qquad
M^{(F)}_{ij}=\left\{\begin{array}{cl}
-\frac{1}{2}\frac{L}{L-1} & |i-j|=1 \\
\mu(-1)^i                 & i=j \\
0                         & \mbox{else}
\end{array}\right. .
\ee
They can be diagonalized and we denote by $\lambda_1 <\lambda_2 < \cdots < \lambda_L$ the eigenvalues of $M^{(F)}$
and by $\omega_1^2, \cdots, \omega_L^2$ those of $M^{(B)}$. The ground state factorizes in the tensor product of a bosonic
and a fermionic ground state. Their energies are given respectively by 
\be
E_0^{(B)} = \frac{1}{2}\sum_{i=1}^L \omega_i , \qquad
E_0^{(F)} = \sum_{i=1}^{L/2} \lambda_i , 
\ee
where we assumed $L$ to be even. These expressions are simply the ground state of a multi-dimensional quantum harmonic
oscillator and that of a system of free (non-relativistic) fermions at half filling. The bound on the ground state energy density
is 
\be
\rho^{(L)} = \frac{1}{L}(E_0^{(B)} + E_0^{(F)}) .
\ee
It can be evaluated numerically without difficulty for large $L$ and its plot is shown in 
Fig.~(\ref{fig:free}) for various $\mu$. The model with linear prepotential is known to respect
supersymmetry ($\rho=0$). We thus expect that $\rho^{(L)}<0$ for all $L$ and also that $\rho^{(L)}\to 0$ as 
$L\to \infty$ with a leading correction $\sim 1/L$. These expectations are fully in agreement with the figure.

\section{Description of the Simulation Algorithm}
\label{app:worldline}

The world-line path integral computes the expectation value of an operator $A$ 
at inverse temperature $\beta$ as
$\mbox{Tr}( A\ e^{-\beta H})/\mbox{Tr}\ e^{-\beta H}$.
As usual the trace is computed by splitting $e^{-\beta H}$ and inserting with time 
spacing $\varepsilon$ complete sets
of intermediate states that are sampled stochastically. 

In more details, we perform a first splitting separating the bosonic terms and the fermionic ones
(we omit the base site because of spatial translation invariance)
\be
h^{(L)} = h_B^{(L)} + h_F^{(L)}
\ee
where 
\ba
h_B^{(L)} &=& \sum_{n=0}^{L-1} \left(\frac 1 2 p_n^2 + \frac 1 2 V(\varphi_n)^2 + \frac 1 4 \varphi_n^2\right) + \\
&& \frac 1 2 \ \frac{L}{L-1} \sum_{n=0}^{L-2} (V(\varphi_n)\varphi_{n+1}-\varphi_n V(\varphi_{n+1})) + \\
&& -\frac 1 4  \ \frac{L}{L-2} \sum_{n=0}^{L-3} \varphi_n\varphi_{n+2}
\ea
and the remaining fermionic terms define $h_F^{(L)}$  as
\ba
h_{F,n}^{(L)} &=& -\frac 1 2 \frac{L}{L-1}(c^\dagger_n c_{n+1} + h.c.) + 
+ \zeta_n (-1)^n V'(\varphi_n) \left(N_n-\frac 1 2 \right)  \nonumber\\
&& + \zeta_{n+1} (-1)^{n+1} V'(\varphi_{n+1}) \left(N_{n+1}-\frac 1 2 \right),
\ea
where $\zeta_n$ is $1$ on sites $0$ and $L-1$ and $1/2$ elsewhere. 

\begin{center}
\begin{figure}[htb]
\epsfig{width=10cm,angle=0,file=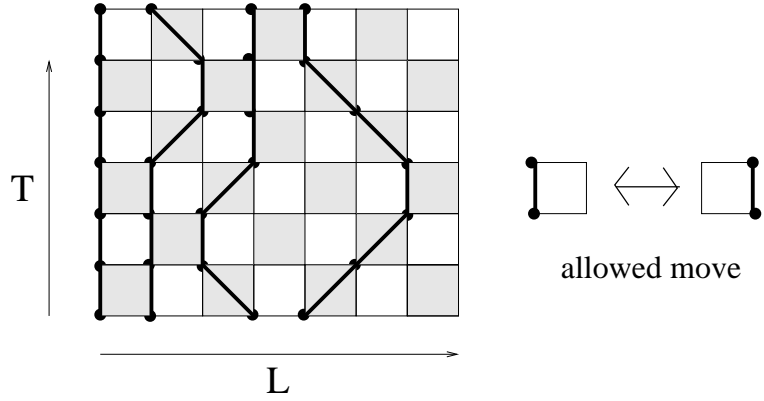}
\caption{Space time lattice. The shaded plaquettes have $(x+t)\mathop{\rm mod} 2 = 0$. 
The boundary conditions are open in the space and closed in time.}
\label{fig:grid}
\end{figure}
\end{center}

The calculation of $\mbox{Tr} e^{-\beta h}$ is written as usual 
\be
\mbox{Tr} e^{-\beta h} = \lim_{\varepsilon\to 0} \sum_{\psi^{(0)}, \psi^{(1)}, \dots, \psi^{(T-1)}}
\langle\psi^{(0)} | e^{-\varepsilon h_B}e^{-\varepsilon h_F}|\psi^{(T-1)}\rangle\cdots
\langle\psi^{(1)} | e^{-\varepsilon h_B}e^{-\varepsilon h_F}|\psi^{(0)}\rangle
\ee
where $T\varepsilon \equiv \beta$ and $|\psi^{(k)}\rangle$ denotes a complete set of states at the $k$th time slice.
We work in a basis of states where the bosonic field $\varphi$ and the fermion occupation numbers are 
diagonal and write therefore
\be
|\psi^{(k)}\rangle = |\varphi^{(k)}, n^{(k)}\rangle
\ee
where $\varphi^{(k)}$, $n^{(k)}$ are the eigenvalues of the above operators (vectors of $L$ numbers, one for
each site). The trace can be written in the following form 
\be
\mbox{Tr}\ e^{-\beta h} = \lim_{\varepsilon\to 0} 
\sum_{n^{(0)}, n^{(1)}, \dots, n^{(T-1)}} 
\int\prod_{t=0}^{T-1}{\cal D}\varphi^{(t)}\ 
e^{-S_B(\varphi)+\log W(\varphi, n)} ,
\ee
where the action $S_B(\varphi)$ is
\ba
S_B &=& \sum_{n,t} \frac{1}{2\varepsilon}(\varphi_{n,t+1}-\varphi_{n,t})^2 + \varepsilon \sum_t {\cal V}_t \nonumber \\
{\cal V}_t &=& \sum_{n}\left(  \frac 1 2 V(\varphi_{n,t})^2 
+ \frac 1 4 \varphi_{n,t}^2 +
\frac 1 2 \ \frac{L}{L-1} \sum_{n=0}^{L-2} (V(\varphi_n)\varphi_{n+1}-\varphi_n V(\varphi_{n+1})) + \right. \nonumber \\
&& \left. -\frac 1 4  \ \frac{L}{L-2} \sum_{n=0}^{L-3} \varphi_n\varphi_{n+2} \right)
\ea
and the weight $W$ is the product 
\be
W(\varphi, n) = \langle n^{(0)} | e^{-\varepsilon h_F}|n^{(T-1)}\rangle\cdots
\langle n^{(1)} | e^{-\varepsilon h_F}|n^{(0)}\rangle .
\ee
In order to build a definite sampling procedure, it is convenient to elaborate further the
explicit expression of $W$. This can be achieved by remarking that the Hamiltonian $h_F$ can be further split as
\be
h_F^{(L)} = \sum_{n\ \rm even} h_{F,n}^{(L)}+\sum_{n\ \rm odd} h_{F,n}^{(L)} .
\ee
The even terms commute among themselves as the odd ones do. It is then convenient to put fermions 
on a square lattice of plaquettes as shown in Fig.~\ref{fig:grid} where the rows denote alternatively
evolution with respect to the even/odd terms. The weight $W$ is obtained by multiplying 
the matrix element of $\exp(-\varepsilon h_{F})$ for each shaded plaquette with 
lower-left corner satisfying $(x+t)\mathop{\rm mod} 2 = 0$.

If we denote the initial state of such a plaquette by $|N_n, N_{n+1}\rangle$ and define 
$t=L/(L-1)$, $a = \zeta_n (-1)^n V'(\varphi_n)$, $b = \zeta_{n+1} (-1)^{n+1} V'(\varphi_{n+1})$,
we have the basic results
\ba
e^{-\varepsilon H_{F,n}} |0,0\rangle &=& e^{\frac{\varepsilon} 2 (a+b)} |0,0\rangle ,\qquad
e^{-\varepsilon H_{F,n}} |1,1\rangle = e^{-\frac{\varepsilon} 2 (a+b)} |0,0\rangle ,\nonumber \\
e^{-\varepsilon H_{F,n}} |1,0\rangle &=& (\cosh\varepsilon\rho-\frac r \rho \sinh \varepsilon\rho) |1,0\rangle 
+\frac{t}{2\rho} \sinh\varepsilon\rho |0,1\rangle ,\nonumber \\
e^{-\varepsilon H_{F,n}} |0,1\rangle &=& (\cosh\varepsilon\rho+\frac r \rho \sinh \varepsilon\rho) |0,1\rangle 
+\frac{t}{2\rho} \sinh\varepsilon\rho |1,0\rangle ,\nonumber
\ea
where
\be
t = \frac{L}{L-1},\qquad r=\frac{a-b}{2},\qquad \rho = \frac 1 2 \sqrt{4r^2+t^2}
\ee
The full weight $W$ is obtained by multiplying the suitable coefficient on the r.h.s.\ (depending on the
final state on the plaquette) over all plaquettes~\footnote{Notice that $W>0$ because 
$t>0$, $\rho>0$ and also 
\be
\cosh\varepsilon\rho\pm \frac r \rho \sinh \varepsilon\rho = \frac 1 2 \left(1+\frac r \rho\right) e^{\varepsilon\rho}
+ \frac 1 2 \left(1-\frac r \rho\right) e^{-\varepsilon\rho} > 0
\ee
due to $|r/\rho| < 1$.
}.

After this preliminary discussion of the explicit form of the lattice partition function, we turn to 
the sampling algorithm and also discuss how energy is measured.

\subsection{Monte Carlo sampling}

The configuration update is obtained by first updating the bosonic fields in the fixed fermion background and then 
updating fermions with fixed bosonic fields. The bosonic update is a multi-hit Metropolis sweep.
The fermionic update is an accept-reject step performed on admissible local changes of the 
fermion world-lines. These are described by the transformation 
$|0,1\rangle \to |0,1\rangle$ into $|1,0\rangle \to |1,0\rangle$ and vice versa on all the plaquettes 
with $(x+t)\mathop{\rm mod} 2 = 1$.
A very clean detailed discussion of such algorithms in the context of models with phonon-electron
coupling can be found in Ref.~\cite{WLPI} (see also Ref.~\cite{TopCharge} for the specific case of the Wess-Zumino model).

\subsection{Energy measurement}

Given the set of sampled configurations, we now explain how to measure the average energy, i.e., the
quantity $\mbox{Tr}( H\ e^{-\beta H})/\mbox{Tr}\ e^{-\beta H}$ that we determine,  of course, 
{\em at the discretized level}. For notational convenience we shall denote it by 
$\langle H\rangle_\beta = \langle h_B\rangle_\beta+ \langle h_F\rangle_\beta$.

The calculation of $\langle h_B\rangle_\beta$ is trivial apart from the $p^2$ part. This term is evaluated
by the virial theorem 
\be
\lim_{\beta\to\infty}\langle p_n^2\rangle_\beta = \lim_{\beta\to\infty}\left\langle \varphi_n\frac{\partial h}{\partial\varphi_n}
\right\rangle_\beta .
\ee
The r.h.s.\ does not present any problem because the fermionic terms appearing in $\partial h/\partial\varphi_n$
are diagonal in the chosen occupation numbers basis.

The evaluation of $\langle h_F\rangle_\beta$ can be reduced to the calculation, 
for a given configuration, of the following basic matrix elements~\cite{WLPI}
\be
\frac{\langle f | e^{-\varepsilon h_F} h_F | i\rangle}{\langle f | e^{-\varepsilon h_F} | i\rangle},\qquad
\frac{\langle f | h_F e^{-\varepsilon h_F} | i\rangle}{\langle f | e^{-\varepsilon h_F} | i\rangle} ,
\ee
where $|i\rangle$ and $|f\rangle$ can be $|00\rangle$, $|11\rangle$, $|01\rangle$, $|10\rangle$.
For transitions without hopping these read
\be
\frac{\langle n, n | e^{-\varepsilon h_F} h_F | n, n\rangle}{\langle n, n | e^{-\varepsilon h_F} | n, n\rangle} = 
\frac{\langle n, n | h_F e^{-\varepsilon h_F} | n, n\rangle}{\langle n, n | e^{-\varepsilon h_F} | n, n\rangle} = 
(a+b)\left(n-\frac 1 2 \right) ,
\ee
and for  transitions with hopping
\ba
\frac{\langle 1, 0  | e^{-\varepsilon h_F} h_F | 1, 0 \rangle}{\langle 1, 0  | e^{-\varepsilon h_F} | 1, 0 \rangle} &=& 
\frac{a-b}{2}-\frac{t}{2}\frac{\cmixed}{\cminus}, \nonumber \\
\frac{\langle 0, 1 | e^{-\varepsilon h_F} h_F | 1, 0 \rangle}{\langle 0, 1  | e^{-\varepsilon h_F} | 1, 0 \rangle} &=& 
\frac{-a+b}{2}-\frac{t}{2}\frac{\cminus}{\cmixed}, \nonumber \\
\frac{\langle 0, 1 | e^{-\varepsilon h_F} h_F | 0, 1\rangle}{\langle 0, 1  | e^{-\varepsilon h_F} | 0, 1\rangle} &=& 
\frac{-a+b}{2}-\frac{t}{2}\frac{\cmixed}{\cplus}, \nonumber \\
\frac{\langle 1, 0 | e^{-\varepsilon h_F} h_F | 0, 1\rangle}{\langle 1, 0  | e^{-\varepsilon h_F} | 0, 1 \rangle} &=& 
\frac{a-b}{2}-\frac{t}{2}\frac{\cplus}{\cmixed}, \nonumber 
\ea
and
\ba
\frac{\langle 1, 0  | h_F e^{-\varepsilon h_F}  | 1, 0 \rangle}{\langle 1, 0  | e^{-\varepsilon h_F} | 1, 0 \rangle} &=& 
\frac{a-b}{2}-\frac{t}{2}\frac{\cmixed}{\cminus}, \nonumber \\
\frac{\langle 0, 1 | h_F e^{-\varepsilon h_F}  | 1, 0 \rangle}{\langle 0, 1  | e^{-\varepsilon h_F} | 1, 0 \rangle} &=& 
\frac{a-b}{2}-\frac{t}{2}\frac{\cplus}{\cmixed}, \nonumber \\
\frac{\langle 0, 1 | h_F e^{-\varepsilon h_F}  | 0, 1\rangle}{\langle 0, 1  | e^{-\varepsilon h_F} | 0, 1\rangle} &=& 
\frac{-a+b}{2}-\frac{t}{2}\frac{\cmixed}{\cplus}, \nonumber \\
\frac{\langle 1, 0 | h_F e^{-\varepsilon h_F}  | 0, 1\rangle}{\langle 1, 0  | e^{-\varepsilon h_F } | 0, 1 \rangle} &=& 
\frac{-a+b}{2}-\frac{t}{2}\frac{\cminus}{\cmixed} . \nonumber 
\ea
where
\be
k_0 = \frac{t}{2\rho} \sin\varepsilon\rho,\quad
k_\pm = \cosh\varepsilon\rho \pm \frac r \rho\sinh\varepsilon\rho
\ee

\end{document}